\documentclass[aps,prb,twocolumn,showpacs,superscriptaddress]{revtex4-1}

\usepackage{graphicx}
\usepackage{dcolumn}
\usepackage{bm}
\usepackage{amsmath} 
\usepackage[dvips]{color}

\newcommand{\EqLabel}[1]{\label{#1}}

\begin{document}

\title{Trapping of three-dimensional Holstein polarons by various impurities}

\author{Hadi Ebrahimnejad}
\affiliation{Department of Physics and Astronomy, University of British
  Columbia, Vancouver, BC, Canada, V6T 1Z1} 

\author{Mona Berciu}
\affiliation{Department of Physics and Astronomy, University of British
  Columbia, Vancouver, BC, Canada, V6T 1Z1} 
\affiliation{Quantum Matter Institute, University of British
  Columbia, Vancouver, BC, Canada, V6T 1Z4} 

\date{\today}

\begin{abstract}
We study the bound states of a three-dimensional Holstein polaron near
various kinds of single impurities, using the momentum average
approximation. We show that the electron-phonon coupling is
responsible for a strong renormalization of the impurity potential,
resulting in an effective potential with significant retardation
effects, which describes essential physics ignored by "instantaneous"
approximations. The accuracy of our approximation is gauged by
comparison with results from Diagrammatic Monte Carlo for the case of
an impurity that modifies the on-site energy of the electron. We also
discuss impurities that modify the local strength of the
electron-phonon coupling, as well as isotope substitutions that
change both the electron-phonon coupling and the phonon frequency, and contrast and
highlight the difference between these cases.
\end{abstract}

\pacs{71.38.-k,71.23.An,72.10.Di}
\maketitle

\section{Introduction}

The challenge to understand the effects of disorder on the behavior of
particles strongly coupled to bosons from their environment is
commonly encountered in correlated electron systems. For example,
high-T$_c$ cuprates are doped anti-ferromagnetic insulators in which,
beside strong coupling to magnons, ARPES measurements have also
provided evidence for strong electron-phonon
coupling.\cite{elphcuprates} At the same time, charge carriers moving
in the CuO$_2$ layers are subject to random disorder potentials from
the adjacent dopant layers. Substituting only a few percent of the Cu
atoms with impurities suppresses superconductivity by localizing the
low energy electronic states.\cite{Basov} Inhomogeneities in the
superconducting gap measured in high resolution tunnelling experiments
have been attributed to atomic scale disorder in the phonon energy and
the electron-phonon coupling strength in these materials.\cite{stm}
Organic semiconductors are another class of materials where interplay
between disorder and electron-phonon coupling is believed to be
important in determining their properties, and are currently under
active investigation.\cite{mish,ciuchi}

Although the results we present here are valid for any type of bosons
(so long as they can be modelled as dispersionless Einstein
modes), for simplicity in the following we restrict our discussion to
optical phonons.  The result of the interplay between disorder and
coupling to such phonons depends on their relative strengths. Disorder
that is considered weak for free electrons can be strong
enough to localize a polaron, that is the dressed quasiparticle which
consists of the electron together with its cloud of phonons, because
of its heavy effective mass. On the other hand, whereas in the weak
disorder regime electron-phonon coupling hinders the motion of
electrons, such coupling actually facilitates the electron mobility in
the strongly disordered regime where the Anderson localization
prevails.\cite{girvin}

Certain aspects of this problem have been studied with various
approximations, most of which rely on sophisticated non-perturbative
methods \cite{nonpert} such as the statistical dynamic mean field theory
(DMFT),\cite{fehske,ciuchi1} or dynamical coherent potential
approximation (DCPA).\cite{sumi,bishop} The underlying meaning of
these approximations and their accuracy is rather hard to gauge.
On the computational side, refined versions of the approximation-free
diagrammatic Monte Carlo (DMC) technique\cite{mishchenko} and of the
continuous quantum Monte Carlo algorithm\cite{hague} have recently
been applied to the problem of a Holstein polaron near a single
impurity. While essentially exact, such calculations require
significant computational resources, and cannot be easily generalized
to other couplings, for example.

Here, we study the bound state formation for a three-dimensional
Holstein polaron in the presence of an impurity, using a
generalization of the Momentum Average (MA) approximation to
inhomogeneous systems.\cite{EPL} MA was originally developed to study
homogeneous systems with various types of electron-phonon
coupling.\cite{MA0,glen,dominic} It is a non-perturbative method that
sums in a closed-form expression all the self-energy diagrams  up to exponentially
small corrections that are neglected. The method can be systematically
improved,\cite{MA1} therefore providing a fast yet accurate way to
scan a vast range of parameters. The method can be used to study all
possible types of disorder for various types of electron-phonon
coupling. Here, we use it to consider different types of disorder due
to single impurities, namely a variation in the on-site energy, in the
electron-phonon coupling and/or in the phonon energy are separately
considered. The accuracy of this method is demonstrated for the former
case by comparison with available DMC results.

Unlike most other theoretical approximations, MA has the important
benefit that its structure reveals the essential physics of such
problems. It is well-known that electron-phonon coupling leads to the
dressing of the particle, resulting in a polaron with a
larger effective mass. What MA reveals is that the electron-phonon
coupling is also responsible for a {\em renormalization of the
disorder potential}. This renormalization can be very large and has
strong retardation effects. Moreover, the renormalized potential can
have a finite-range even if the bare disorder is on-site only. The
single impurity problem provides us with a simple test case to
understand the effects of this renormalization, and to accurately
compare and contrast the behavior of the polaron in the presence of
various types of local disorder. Such results are a necessary first
step in order to gain the intuition needed for understanding the
behavior of more complicated systems.

The paper is organized as follows: In Section II we describe our
method and discuss its meaning (full details are 
provided in the Appendix). Section III presents our results for the
three types of impurities, and Section IV contains the summary.

\section{Momentum average approximation for inhomogeneous systems}

For completeness, we present the MA formalism for the general case
of random on-site disorder plus inhomogeneities in both the
coupling and the phonon frequencies. A simpler case (with only on-site
disorder) has been briefly
discussed in Ref. \onlinecite{EPL}. The  Hamiltonian is:
\begin{equation}
\label{1}
  {\cal H} = {\cal H}_d + \hat{V}_{el-ph}={\cal H}_0+ \hat{V}_d + \hat{V}_{el-ph},
\end{equation}
where ${\cal H}_d$ describes the non-interacting part of the
Hamiltonian, and for convenience is further divided into:
$$
{\cal H}_0 = -t\sum_{\langle i,j\rangle}(c^{\dagger}_ic_j+h.c.) +
\sum_i  \Omega_i {b^{\dagger}_i} b_i
$$
which contains the
kinetic energy of the particle and the energy of the  boson
modes $(\hbar=1)$, plus 
$$
\hat{V}_d =\sum_i \epsilon_i {c^{\dagger}_i} c_i
$$
describing the on-site  disorder. The interaction part:
$$
\hat{V}_{el-ph} =\sum_i g_i {c^{\dagger}_i} c_i (b^{\dagger}_i + b_i)
$$
describes the (possibly inhomogeneous) Holstein-like
coupling\cite{Holstein} between the particle and the bosons. Here, $i$
indexes lattice sites -- the lattice can be in any dimension, and of
finite or infinite extent. The operators $c_i$ and $b_i$ describe,
respectively, particle and boson annihilation from the corresponding
state associated with lattice site $i$. The spin of the particle is
ignored because it is irrelevant in this case, but generalizations are
straightforward.\cite{Luci} For simplicity, we assume nearest-neighbour
hopping; this approximation can also be trivially relaxed. Depending
on the model of interest, the on-site energies $\epsilon_i$,
electron-phonon couplings $g_i$ and phonon frequencies $\Omega_i$ can
be assumed to be random variables. As detailed below, our results here
will focus on single on-site impurities such as $\epsilon_i = -U
\delta_{i,0}$, but the formalism applies for any model consistent with
Eq. (\ref{1}).

Our goal is to calculate the single polaron Green's function in real space:
\begin{equation}
G_{i j}(\omega) = \langle 0| c_i \hat{G}(\omega) c^{\dagger}_j | 0
  \rangle = \sum_{\alpha} \frac{\langle 0| c_i | \alpha \rangle
	\langle \alpha | c^{\dagger}_j | 0
	\rangle}{\omega-E_\alpha+i\eta},  
\end{equation} 
where $|0\rangle$ is the vacuum,
$\hat{G}(\omega)=[\omega-{\cal H}+i\eta]^{-1}$ is the resolvent with $\eta \rightarrow 0_+$,
and $E_{\alpha}$,
$|\alpha \rangle$ are single polaron eigenenergies and eigenstates of
the Hamiltonian, ${\cal H}|\alpha\rangle = E_\alpha
|\alpha\rangle$. Knowledge of this Green's function allows us to find
the spectrum from the poles, and the
local density of states (LDOS) measured in scanning tunneling
microscopy (STM) experiments,
$\rho(i;\omega)=-\frac{1}{\pi} \mbox{Im} G_{ii}(\omega)$.

To calculate this quantity, we use repeatedly Dyson's identity:
$\hat{G}(\omega)=\hat{G}^d(\omega)+\hat{G}(\omega)\hat{V}_{el-ph}\hat{G}^d(\omega)$,
where $\hat{G}^d(\omega)=[\omega-{\cal H}_d+i\eta]^{-1}$ is the resolvent
for the non-interacting system. The first equation of motion
generated this way reads:
\begin{equation} 
\label{eq:Dyson}          
G_{ij}(\omega) = G^d_{ij}(\omega) + \sum_l g_l F^{(1)}_{il}(\omega) G^d_{lj}(\omega).
\end{equation}
where
\begin{equation}
\EqLabel{dis}
G^d_{ij}(\omega)= \langle 0| c_i \hat{G}_d(\omega) c^{\dagger}_j | 0
  \rangle 
\end{equation}
are, in principle, known quantities  and we have introduced the
generalized Green's functions: 
\begin{equation}
\nonumber 
F_{ij}^{(n)}(\omega)= \langle 0| c_i \hat{G}(\omega) c^{\dagger}_j
b^{\dagger n}_j |0\rangle. 
\end{equation}
Note that $F_{ij}^{(0)}(\omega)=G_{ij}(\omega)$. Next, we generate
equations of motion for these generalized Green's functions. For any
$n\ge 1$, we find:
\begin{multline}
F^{(n)}_{ij}(\omega) = \sum_{l\neq j} g_l G^d_{lj}(\omega-n\Omega_j)
\langle 0| c_i G(\omega) c^{\dagger}_l b^{\dagger}_l b^{\dagger n}_j |
0\rangle 	\\
+ g_j	G^d_{jj}(\omega-n\Omega_j)\left[F_{ij}^{(n+1)}(\omega)+ 
nF_{ij}^{(n-1)}(\omega)\right] \EqLabel{Fex}   
\end{multline}

This equation relates $F^{(n)}$ not only to other Green's
functions of a similar type, but  also introduces new propagators with
phonons at two different sites. Equations of motions can be
calculated for these new generalized Green's functions, linking them
to yet more general Green's functions, and so on and so forth. The
resulting hierarchy of coupled equations describes the problem
exactly, but is  unmanageable. Approximations are needed to simplify
it and find a closed-form solution.

The main idea behind the MA approximations is to simplify this set of
equations by neglecting exponentially small contributions in each
equation of motion. At the simplest level -- the so-called MA$^{(0)}$
approximation --  we ignore the first term in Eq. (\ref{Fex})  for
any $n\ge 1$. This is reasonable at low-energies like the ground state
(GS), $\omega\sim E_{GS}$, where $\omega-n\Omega_j$ is well below the
energy spectrum of ${\cal H}_d$ and, therefore,
$G^d_{lj}(\omega-n\Omega_j)$ is guaranteed to decrease exponentially with increasing
distance $|l-j|$. As a result, here we keep the largest $l=j$
propagator, and ignore exponentially smaller $l\ne j$
contributions. Although this is the simplest  possible such
approximation, it is
already accurate at low energies, as shown in the results section. It can also be
systematically improved, as discussed below. First, however, we
complete this  MA$^{(0)}$-level solution.

\begin{figure}[t]
\includegraphics[width=0.9\columnwidth]{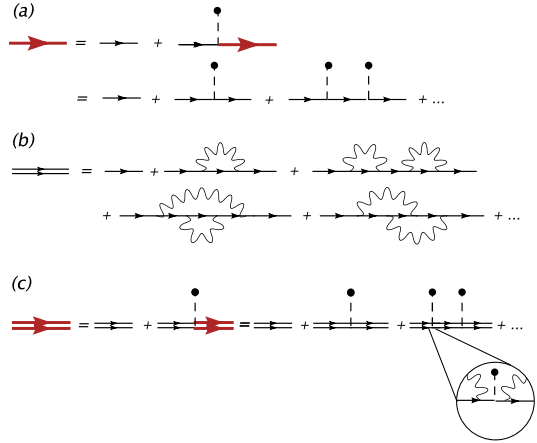}
\caption{(color online) 
 Diagrammatic expansion for (a) the  disorder Green's function
 $G^d_{ij}(\omega)$  (bold red line); (b) the polaron Green's function in a
 clean system   $G^{(0)}_{ij}(\tilde{\omega})$ (double thin line), and
 (c) the ``instantaneous'' approximation for the polaron Green's function in a
 disordered system, $G^{d}_{ij}(\tilde{\omega})$ (double bold red
 line). The thin black lines depict free electron propagators, the
 wriggly lines correspond to phonons, and scattering on
  the disorder potential is depicted the dashed lines ending with circles. See text
 for more details.} 
\label{fig1}
\end{figure}

The simplified equation of motion now reads:
\begin{equation}
\label{eq:recursive}
\nonumber
	F_{ij}^{(n)}(\omega)=g_jG^d_{jj}(\omega-n\Omega_j)\left[F_{ij}^{(n+1)}(\omega)
	+nF_{ij}^{(n-1)}(\omega)\right].
\end{equation}  
On physical grounds, we know that $F^{(n)}_{ij}(\omega)$ must vanish for
sufficiently large $n$, because its Fourier transform  is
the amplitude of probability that a particle injected in the system
will generate $n$ phonons in time $t$, and this must vanish for large
enough $n$. As a result, these recursive equations admit the solution:
$$F^{(n)}_{ij}(\omega) = A_n(j,\omega) F_{ij}^{(n-1)}(\omega),$$ where the
continued-fraction:
\begin{equation}
  \label{An}
  	A_n(j,\omega)=\frac{ng_jG^d_{jj}(\omega-n\Omega_j)}{1-g_jG^d_{jj}(\omega-n\Omega_j)
  	A_{n+1}(j,\omega)} 
\end{equation}  
can be efficiently evaluated starting from a cutoff $A_{N_c}(j,\omega)=0$
for a sufficiently large ${N_c}$. Generally speaking, this cutoff $N_c$ must
be much larger than the average number of phonons expected at site
$j$; in practice, the cutoff is increased until convergence is reached to
within the desired accuracy. Substituting $F^{(1)}_{ij}(\omega) =
A_1(j,\omega) G_{ij}(\omega)$ in Eq.~(\ref{eq:Dyson}) 
leads to a closed system of linear
equations for the original Green's function:
\begin{equation}  
\label{eq:G-G-Gd}        
G_{ij}(\omega) = G^d_{ij}(\omega) + \sum_l G_{il}(\omega)g_l 
A_1(l,\omega)G^d_{lj}(\omega). 
\end{equation}
This equation has a  similar structure to the equation linking the disorder Green's
function to the free particle propagator 
$G^{(0)}_{ij}(\omega)=\langle
0 | c_i 
[\omega + i \eta - {\cal H}_0]^{-1} c_j^\dagger|0\rangle$ (in the
absence of coupling to phonons), which is depicted diagrammatically in
Fig. \ref{fig1}(a), and which reads:
\begin{equation}
\EqLabel{clean}
G^d_{ij}(\omega) = G^{(0)}_{ij}(\omega) + \sum_l 
G^d_{il}(\omega) \epsilon_l G^{(0)}_{lj}(\omega). 
\end{equation}
This analogy shows that coupling to phonons
renormalizes the on-site disorder   $\epsilon_l \rightarrow \epsilon_l +
g_l A_1(l,\omega)$. Note that $A_1(l,\omega)$ depends not only on the
local phonon frequency $\Omega_l$ and coupling $g_l$, but also on {\em
  all} the bare on-site energies $\epsilon_i$ through the disorder propagators
$G^d$. This is the simplest example of the emergence of a renormalized potential
for this problem, that is made very transparent within the MA approximation.

While Eq. (\ref{eq:G-G-Gd}) can be solved directly for a
finite-size system, we can improve its efficiency and reveal a
different physical interpretation by explicitly
removing the ``average'' contribution due to the electron-phonon
interactions. 
Let $g$ and $\Omega$ be the average
values of the $g_i, \Omega_i$ distributions. We assume that the
on-site energy average $\epsilon=0$ (a finite value 
results in a trivial shift of all energies). Then, let:
\begin{equation}
\EqLabel{Ac}
  	A_n(\omega)=\frac{ngg_0(\omega-n\Omega)}{1-g
  	g_0(\omega-n\Omega)   	A_{n+1}(\omega)} 
\end{equation}
be the continued fractions corresponding to these average parameters,
where we use the short-hand 
notation:
$$
g_0(\omega) = G^{(0)}_{ii}(\omega) = {1\over N} \sum_{\bf{k}}^{} {1\over
  \omega - \epsilon_{\bf k} + i\eta}
$$
for the on-site free propagator  (in the absence of disorder, this quantity
becomes independent of the site). It is given by the momentum average
of the 
free propagator, where $\epsilon_k$ is the
free-particle dispersion.

The ``average'' renormalization of
the on-site energy is now recognized to represent the corresponding
MA$^{(0)}$ self-energy for a ``clean'' system, i.e. a homogeneous system with  average coupling and
phonon frequency:
$$
\Sigma_\mathrm{MA^{(0)}}(\omega) = g A_1(\omega)
$$
see, for instance, Eqs. (11) and (12) of Ref. \onlinecite{MA1}.

Introducing the effective  disorder potential:
\begin{equation}
\label{v0}
v_0(l,\omega)=g_lA_1(l,\omega)-\Sigma_\mathrm{MA^{(0)}}(\omega),
\end{equation}
Eq. (\ref{eq:G-G-Gd})  can be rewritten as:     
\begin{equation} 
\label{eq:G-v0}         
G_{ij}(\omega) = G^d_{ij}(\tilde{\omega}) + \sum_l G_{il}(\omega)
v_0(l,\omega)G^d_{lj}(\tilde{\omega}), 
\end{equation}
where $\tilde{\omega}=\omega-\Sigma_\mathrm{MA^{(0)}}(\omega)$. This energy
renormalization, $\omega \rightarrow \tilde{\omega}$, reflects the
fact that processes describing the formation of the polaron
in the ``clean'' system   have been explicitly summed. 

Besides being numerically more efficient, since now $v_0(j,\omega)$
contains only the fluctuations from the (not necessarily small) average
value included in $\tilde{\omega}$, Eq. (\ref{v0}) reveals a
different interpretation for the effects of the interplay between
disorder and electron-phonon coupling.

Consider first the meaning of 
$G^d_{ij}(\tilde{\omega})$, which would be the solution if we
could ignore $v_0(j,\omega)$. In the absence of on-site disorder
 this term equals
$G^{(0)}_{ij}(\tilde{\omega})$, i.e. the expected
solution for a polaron in the clean system, depicted diagrammatically in
Fig. \ref{fig1}(b) (of course, the exact self-energy is here
approximated by  $\Sigma_\mathrm{MA^{(0)}}(\omega)$). Comparing
Figs. \ref{fig1}(a) and \ref{fig1}(b), it follows that
$G^d_{ij}(\tilde{\omega})$ is the sum of the diagrams shown in
Fig. \ref{fig1}(c).

At first sight, this seems to be the full answer for this problem, since
these diagrams sum the contributions of all the processes in which the
polaron scatters once, twice, etc., on the disorder potential. This is
certainly the answer obtained in the limit of an ``instantaneous''
approximation valid when $\Omega\rightarrow \infty$, i.e. when
the ions are very light and respond instantaneously to the motion of
electrons. In this case, one can perform a Lang-Firsov transformation
and after an additional averaging over phonons, one obtains an
approximative effective 
Hamiltonian:\cite{hague,inst} 
\begin{equation}
\EqLabel{inst}
{\cal H}_{inst} = -t^*\sum_{\langle i,j\rangle}^{} (c_i^\dagger c_j +
h.c.) +\sum_{i}^{} \left(\epsilon_i - {g^2\over \Omega} \right)c_i^\dagger c_i
\end{equation}
where $t^* = t e^{-g^2\over \Omega^2}$ is the renormalized polaron hopping,
and $-g^2/\Omega$ is the polaron formation energy (for simplicity,
here we
assume that the phonon energies and electron-phonon coupling are
homogeneous, $g_i \rightarrow g, \Omega_i\rightarrow
\Omega)$. The Green's function of this Hamiltonian
is also given by Fig. \ref{fig1}(c), if the polaron propagator
is approximated by:
$$G^{(0)}_{ij}(\tilde{\omega}) \rightarrow {1\over N} \sum_{\bf{k}}^{}
\frac{e^{i \bf{k}\cdot(\bf{R}_i-\bf{R}_j)}}{\omega- \epsilon^*_{\bf k} +
  {g^2\over \Omega} + i\eta}$$
where $\epsilon^*_{\bf k}$ is the renormalized kinetic energy. Of
course, using the full expression of $G^{(0)}_{ij}(\tilde{\omega})$ is
preferable since the self-energy
$\Sigma_\mathrm{MA^{(0)}}(\omega)$ describes much more accurately the overall
energy shift and effective mass renormalization than those asymptotic
expressions, besides also including the quasiparticle weight. 

That $G^d_{ij}(\tilde{\omega})$ cannot be the full answer becomes
obvious if we consider what happens when we rewrite the clean polaron
propagators in terms of free particle and phonon lines, i.e. we
substitute the expansion of  Fig. \ref{fig1}(b) in
\ref{fig1}(c). Doing so reveals that within this ``instantaneous''
approximation, scattering of the electron on the disorder potential is
allowed only when no phonons are present, see zoom-in in
Fig. \ref{fig1}(c). However, we know that for moderate and large
electron-phonon coupling, the probability to find no phonons in the
system is exponentially small, therefore the processes summed in
Fig. \ref{fig1}(c) have very low probabilities. 

\begin{figure}[t]
\includegraphics[width=0.9\columnwidth]{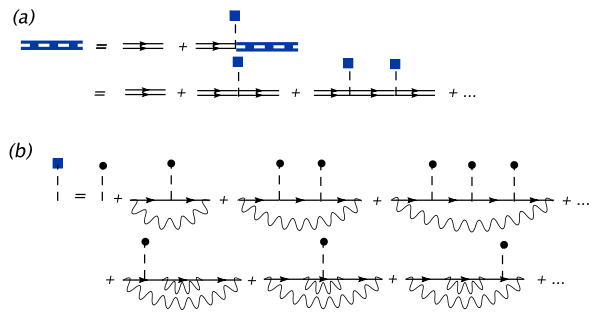}
\caption{(color online) 
 (a) Diagrammatic expansion for the full MA solution $G_{ij}(\omega)$
 (thick dashed blue line) in terms of the clean polaron Green's
 function (double thin line) and scattering on the renormalized disorder
 potential $\epsilon^*_l(\omega)$, depicted by dashed lines ended with
 squares. (b) Diagrammatic expansion of $\epsilon^*_l(\omega)$. For
 more details, see text.} 
\label{fig2}
\end{figure} 

What is missing in Fig. \ref{fig1}(c) are diagrams describing the
scattering of the electron on the disorder potential in the presence
of the phonons from the polaron cloud. Their contribution is included
through the renormalized potential $v_0(l,\omega)$ in the second term
of Eq. (\ref{eq:G-v0}). Indeed, the full MA solution shows that the
polaron scatters not on the bare disorder $\epsilon_l$, but on
the renormalized disorder potential
\begin{equation}
\EqLabel{redis}
\epsilon_l^*(\omega)= \epsilon_l + v_0(l,\omega),
\end{equation}
 as depicted in
Fig. \ref{fig2}(a). The diagrammatic expansion of the additional term
$v_0(l,\omega)$, shown in Fig. \ref{fig2}(b), verifies that it indeed
describes the effective 
scattering in the presence of arbitrary numbers of phonons. 

Taken together, the diagrams summed in Fig. \ref{fig2}
represent all possible contributions to the polaron propagator in the
disordered system. The MA$^{(0)}$ approximation consists in discarding
exponentially small contributions from each of these diagram, as already
discussed. MA$^{(0)}$ also has an exact variational meaning, namely of
assuming that the polaron cloud can have phonons only on a single
site, in direct analogy with the  clean case.\cite{MA1,bar} It is quite
remarkable that all the diagrams corresponding to  this variational
approximation can still be summed analytically in closed form, even in
the presence of disorder. 

As is the case for the clean system, MA can be systematically improved
by keeping more contributions to Eq. (\ref{Fex}). In particular, at the
MA$^{(1)}$ level, we also treat the equation for the $F^{(1)}$ functions
exactly, and make the MA approximation of discarding exponentially small
off-diagonal propagators only for $n\ge 2$. The logic here is that the propagators
appearing in   $F^{(1)}$ have the highest energy, therefore the slowest
exponential decay. The MA$^{(1)}$ equations can also be solved in
closed form. The details are presented in the Appendix. The final
solution looks identical to Eq. (\ref{eq:G-v0}), except the
renormalized energy is now $\tilde{\omega} = \omega
-\Sigma_\mathrm{MA^{(1)}}(\omega)$ while the renormalized potential
$v_0(l,\omega)$ is replaced by a more complicated, yet more accurate
expression $v_1(l,\omega)$. The meaning of all these quantities,
however, is the same. 

To summarize, MA reveals that the role of electron-phonon coupling
is two-fold. On one hand, it renormalizes the quasiparticle properties
due to polaron formation, just like in a clean system (as revealed by
the explicit appearance of the ``clean'' system self-energy). However,
this coupling also renormalizes the 
disorder potential experienced by the particle, $\epsilon_l
\rightarrow \epsilon^*_l(\omega)$. As we show next for various
types of disorder, this renormalization is non-trivial in that it
has strong retardation effects, and has
significant consequences.  ``Instantaneous'' approximations completely
ignore this renormalization, and therefore miss essential physics. To
be fair, in practice the ``instantaneous'' approximations are usually
implemented in an improved, variational form\cite{inst,varapro} which
is certain to be much more accurate. However, this leads to the
necessity to calculate the variational parameters through a
self-consistent loop, making the improved version computationally
as complicated as DMFT and DCPA, which also have
self-consistency loops.

In contrast, MA gives a closed analytical expression for all
quantities of interest, in a formulation whose meaning is very
transparent, and whose accuracy can be systematically
improved.

\section{Polaron near a single impurity}

We now apply the general formalism described above to the
simplest type of ``disorder'', namely an otherwise clean 3D simple
cubic lattice with a single impurity. Impurities which modulate the
on-site energy, the strength of the electron-phonon coupling and/or
the phonon frequency are separately considered. We
investigate under what conditions such impurities can trap the polaron.

As a reference case, we first review here briefly the solution in the absence of
electron-phonon coupling. In this case, the impurity can only modulate
the on-site potential, $\epsilon_i = -U \delta_{i,0}$, and the
Hamiltonian reduces to: 
\begin{equation}
\label{h0dis}
	{\cal
	H}_\mathrm{d}=-t\sum_{<ij>}(c^{\dagger}_ic_j+h.c.)-Uc^{\dagger}_0c_0
	={\cal H}_0+\hat{V}_d.
\end{equation}

For this form of $\hat{V}_d$, Eq. (\ref{clean}) reads:
\begin{equation}
	G^d_{ij}(\omega)=G^{(0)}_{ij}(\omega)-UG^d_{i0}(\omega)G^{(0)}_{0j}(\omega),
\end{equation}
and is trivially solved to find:
\begin{equation}
\label{G_d}
	G^d_{ij}(\omega)=G^{(0)}_{ij}(\omega)-U\frac{G^{(0)}_{i0}(\omega)
          G^{(0)}_{0j}(\omega)} 
	{1+UG^{(0)}_{00}(\omega)}.
\end{equation}

Of course, because of translational and time reversal symmetry,
$G^{(0)}_{ij}(\omega)=G^{(0)}_{i-j,0}(\omega)=G^{(0)}_{j-i,0}(\omega)$,
etc.    

The LDOS at the impurity site is then found to be:
$$\rho(0;\omega)=-\frac{1}{\pi}\mbox{Im} G^{d}_{00}(\omega)=
\frac{\rho_0(0;\omega)}{|1+UG^{(0)}_{00}(\omega)|^2},$$ where
$\rho_0(0;\omega)=-\frac{1}{\pi}\mbox{Im} G^{(0)}_{00}(\omega)$ 
is the LDOS in the clean system (equal to the DOS, because of
translational invariance). As a result, a bound state below the
continuum, signalled by a delta-function peak in $\rho(0;\omega)$,
occurs if and only if the denominator of Eq. (\ref{G_d}) vanishes. For a
3D simple cubic lattice this means that an impurity bound state
appears if there is an energy $E<-6t$ such that $\mbox{Re}
G^{(0)}_{00}(E)=-1/U$ (below the continuum the imaginary part of
$G_{00}^{(0)}(E)$ vanishes). This equation can be solved graphically to find
that a bound state appears for any $U \ge U_c= -{1\over \mbox{Re}
  G^{(0)}_{00}(-6t)}\sim 3.96t$.

In the presence of electron-phonon coupling the equations are more
complicated, but the idea is the same: we calculate the LDOS at
the impurity site and compare it to the DOS of the clean system. If
the former has a peak below the threshold of the latter, then a bound
state exists at that energy. We then vary $U$ to find the
critical value above which a bound state is guaranteed. 
More details about the impurity state,
such as its localization length, statistics for the phonon cloud,
etc., can be extracted from the LDOS at
sites in the neighbourhood of the impurity. Here we focus on
identifying when bound impurity states are stable.

\subsection{Impurity changing the on-site energy}

The Hamiltonian describing this case 
is:
\begin{equation}
\label{Ham_single}
	{\cal H}=\hat{T}+\Omega\sum_ib^{\dagger}_ib_i+g\sum_i
	c^{\dagger}_ic_i(b_i+b^{\dagger}_i) -Uc^{\dagger}_0c_0, 
\end{equation}
where $\hat{T}$ is the electron's tight-binding Hamiltonian. We are
interested in the attractive impurities, $U>0$, when an impurity state can be bound
near the impurity site. To find
the LDOS at the impurity site, we need to
solve Eq.  (\ref{eq:G-v0}) to find $G_{00}(\omega)$. Note that now
$G^d_{ij}(\omega)$ is known, being 
given by Eq. (\ref{G_d}). The free-particle propagators
$G^{(0)}_{ij}(\omega)={1\over N} \sum_{\bf{k}}^{} \frac{e^{i
	\bf{k}\cdot(\bf{R}_i-\bf{R}_j)}}{\omega- \epsilon_{\bf k}  
   + i\eta}$, where for the simple cubic lattice $\epsilon_{\bf k}=
-2t \sum_{i=1}^{3}\cos k_i$,  
can be calculated by doing the integrals over the Brillouin zone. A
more efficient approach, which we use, is discussed in Ref. \onlinecite{Ash}.

In this case, the renormalized impurity potential
decays fast at sites away from the impurity, because
$G^{d}_{ll}(\omega) \rightarrow G^{(0)}_{ll}(\omega) $ when 
$|l|\rightarrow \infty$, so that $A_1(l,\omega) \rightarrow
A_1(\omega)$. Physically, this is because the impurity has less and
less influence at sites far from where it is located. Mathematically,
this follows from Eq. (\ref{G_d}) and the fact that
$G^{(0)}_{l0}(\omega)$ decreases exponentially with the distance
between site $l$ and the origin, at energies below the free particle
continuum, of interest here. As a result, in Eq. (\ref{eq:G-v0}) we
only need to sum over sites $l$ close to the impurity, and the system can
be solved very efficiently.

\begin{figure}[t]
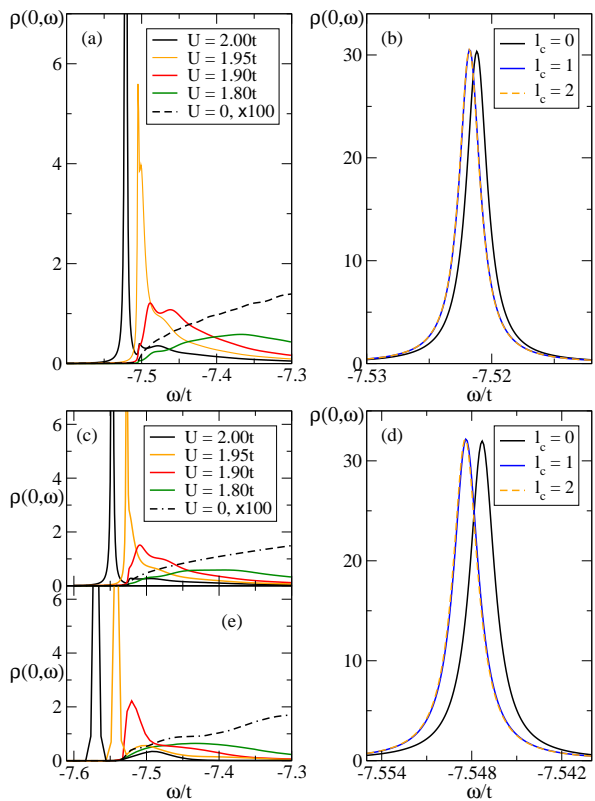

\includegraphics[width=0.9\columnwidth]{fig3a.eps}
\includegraphics[width=0.9\columnwidth]{fig3b.eps}
 \caption{(color online) LDOS at the impurity site $\rho(0,\omega)$ in units of
 $1/t$, vs. the energy
 $\omega/t$, for (a) MA$^{(0)}$ with $l_c=0$ for
 $U=1.8,1.9, 1.95$ and 2.0. The dashed line shows the DOS for the
 clean system, times 100; (b) MA$^{(0)}$ at $U/t=2$, and cutoffs in the
 renormalized potential $l_c = 0,1,2$.  Panels (c) and (d) are the
 same as (a) and (b), 
 respectively, but using MA$^{(1)}$. Panel (e) shows  DMC results from
 Ref. \onlinecite{mishchenko}, for same parameters as (a) and
 (b). Other parameters are 
 $\Omega=2t, \lambda=0.8, \eta/t=10^{-3}$.}  
 \label{fig3}
 \end{figure}

The appropriate value for this cutoff varies depending on the various
parameters, but generically it decreases as the energies of interest
become lower. Again, mathematically this is due to the exponential
decrease of the free particle propagator with distance, and the
fact that this decrease becomes faster as $\omega \rightarrow
-\infty$. Physically, this can be understood as follows. First, let us
explain why is the renormalized disorder potential non-local, even
though the bare impurity potential is local. The answer is provided by
the diagrams which contribute to it, see Fig. \ref{fig2}(b). Consider,
for simplicity, the MA$^{(0)}$ approximation. In this case, all phonon
lines appearing in these diagrams start and end at the site $l$ for which
$v_0(l,\omega)$ is being calculated -- this is the site where the
phonon cloud is located. However, the electron is found with various
probabilities away from the polaron cloud, so it can scatter on the
impurity if this is located within the ``radius'' of the polaron,
where the electron resides. In other words, the range of the
renormalized potential is controlled by the polaron size. From medium
to large couplings, as the polaron becomes smaller, the renormalized
potential becomes more local. At small couplings, though,
the distance between the electron and its phonon cloud can be
appreciable, and the range of the potential increases.

\begin{figure}[t]
\includegraphics[width=0.9\columnwidth]{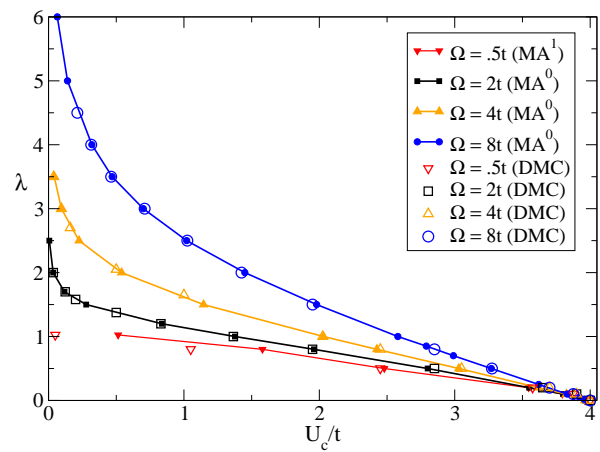}
 \caption{(color online) Phase diagram separating the regime where the polaron is
 mobile (below the line) and trapped (above the line). The effective
 coupling is $\lambda=g^2/(6t\Omega)$, and the critical trapping
 potential $U_c$ is shown for several values of $\Omega/t$. The MA
 results (filled symbols) compare well  with the DMC results of
 Ref. \onlinecite{mishchenko} (empty symbols).} 
 \label{fig4}
 \end{figure}

In Fig. \ref{fig3}(a) we plot $\rho(0,\omega)$ over a wider energy
range, for several values of $U$, within MA$^{(0)}$. The dashed line
shows the DOS of the clean system, multiplied by 100 for
visibility. For $U/t=1.8, 1.9$, the impurity attraction is not
sufficient to bind a state below the continuum, although the LDOS is
pushed towards the lower band-edge. For $U=1.95t$, there is a peak
just below the continuum. Because of the finite value of $\eta$, the
two features are not completely separated and the continuum onset
looks like a ``shoulder'', however lowering $\eta$ allows us to
clearly separate the two features (not shown). Finally, for $U=2t$ the
bound state peak is clearly below the continuum, so 
in this case $U_c\approx1.95t$.  This $U_c$ value equals that obtained in DMC, 
although our energies are overall 
higher than the exact DMC values shown in Fig. \ref{fig3}(e),
as expected for a variational approximation.  At the
MA$^{(1)}$ level the agreement with DMC is significantly improved
since all features move toward lower energies, see
Fig. \ref{fig3}(c). The critical value $U_c\approx 1.95t$ at which an
impurity state appears below the continuum is little affected, however,
by this overall shift of the spectral weights. 

The dependence of the bound state energy on the cutoff is shown in
Figs. \ref{fig3}(b) and (d) for MA$^{(0)}$ and MA$^{(1)}$, respectively, for the case
with $U=2t$, $\Omega=2t$ and the effective coupling $\lambda={g^2\over
  6t\Omega}=0.8$. At these energies, keeping only the local part in MA$^{(0)}$,
i.e. setting $v_0(l,\omega)\rightarrow \delta_{l,0} v_0(l,\omega)$, is already
  a very good approximation. Including the correction from the 6
  nearest neighbor sites ($l_c=1$) lowers the energy somewhat, but
 the contribution from the second nearest neighbor sites ($l_c=2$) is no longer
 visible on this scale for either the MA$^{(0)}$ or the  MA$^{(1)}$ results.

Repeating this process for other values of the parameters, we trace $U_c$
in the parameter space. This is shown in Fig. \ref{fig4}, for
$\Omega/t=0.5,2,4,8$ and various effective couplings.  
The MA results (filled symbols) are in good quantitative agreement 
with the DMC results (empty symbols) for larger $\Omega\ge 1$ values (these
are MA$^{(0)}$ results for cutoff $l_c=0$. Using MA$^{(1)}$ and/or
increasing the cutoff changes the values of $U_c$ by less than 1\%
everywhere we checked). For the smaller frequencies such as
$\Omega=0.5t$, MA is known to become quantitatively less accurate at
intermediary couplings\cite{MA0,MA1}, and indeed, here we see a
discrepancy with the DMC data even for the  MA$^{(1)}$ results. To improve the
quantitative agreement here, one should use a 2- or 3-site MA variational
approximation for the phonon cloud, as discussed in
Refs. \onlinecite{dominic}.

As expected, when $\lambda
\rightarrow 0$, $U_c$ goes towards the expected critical value in the
absence of electron-phonon coupling, of roughly $3.96t$. As the
effective coupling increases $U_c$ decreases, but the lines never
intersect the $y$-axis: $U_c=0$ is impossible, since the polaron
cannot be trapped (localized) in a clean system as long as it has a
finite  effective mass, i.e. for any finite value
of $\lambda$.

The decrease of $U_c$ with increasing $\lambda$ is expected, and is
usually attributed to the fact that the effective polaron mass
increases with $\lambda$, and this makes it easier to
trap near the impurity.\cite{mishchenko} However, we claim that this is not the full
story, and that the renormalization of the trapping potential also
plays a non-trivial role. 

\begin{figure}[t]
\includegraphics[width=0.9\columnwidth]{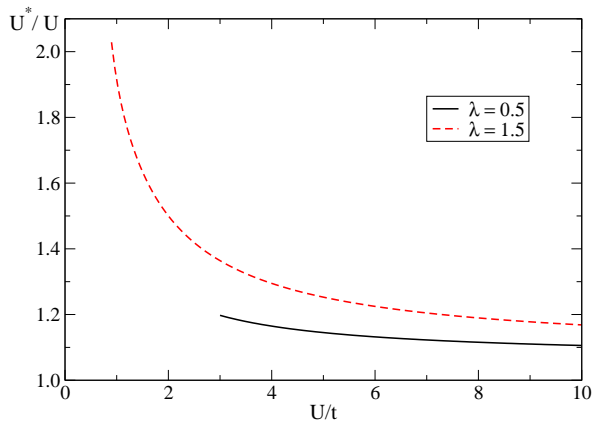}
 \caption{(color online) Effective value of the impurity attraction $U^*/U$,
 extracted from the scaling $E_B^*/t^* = f(U^*/t^*)$, for $\lambda=0.5,
 1.5$ and $\Omega=3t$. In order to have the best fit to the data, we
 plot each curve starting from slightly larger $U/t$ than the
 corresponding $U_c/t$ given in Fig. (\ref{fig4}). For more details,
 see text.} 
 \label{fig5}
 \end{figure}

Consider, first, the Hamiltonian of Eq. (\ref{h0dis}), which describes
the impurity problem in the absence of electron phonon coupling. The
binding energy of the impurity state (once formed) is a monotonic
function of the only dimensionless parameter of this problem:
${E_B\over t}
= f\left( {U\over t}\right)$ for any ${U\over t}\ge {U_c\over
  t}\approx 3.96$. If one views the polaron as a quasiparticle with an
effective hopping $t^*$ which scatters on  the same
potential $U$ as the bare particle (instantaneous approximation), then
the 
polaron binding energy should be ${E^*_B\over t^*}
= f\left( {U\over t^*}\right)$ for any ${U\over t^*}\ge {U_c\over
  t^*}\approx 3.96$. In particular, this predicts $U_c/t = 3.96 t^*/t$
decreasing with increasing $\lambda$, in qualitative agreement with
Fig. \ref{fig4}.

This hypothesis can be tested. The function $f(x)$ is easy to
calculate numerically, we can extract the binding energy $E_B^*$ for
the trapped states by comparing their trapped energy to the GS energy
of the polaron in the clean system, and the effective hopping $t^*\sim
1/m^*$ is directly linked to the effective polaron mass $m^*$ in the
clean system.\cite{MA0} We find that this
scaling is not obeyed. Instead, one needs to also rescale the impurity
potential, i.e. use ${E^*_B\over t^*} = f\left( {U^*\over t^*}\right)$
where $U^*\ne U$. Of course, this scaling assumes that the scattering
potential is local. As we discussed, this is not true
although it is a good approximation for medium and large
couplings. 

In Fig. \ref{fig5} we show the renormalized value $U^*/U$ extracted
this way, as a function of
$U/t$ above the corresponding threshold values $U_c/t$, for a medium and a large
effective coupling $\lambda=0.5, 1.5$ and $\Omega= 3t$. Qualitatively
similar curves are found for other parameters. We see that
$U^*\rightarrow U$ only when $U\rightarrow \infty$ and becomes the
dominant energy scale (hence anything else is a small
perturbation). For fixed $U, \Omega$, we find that $U^*$ increases
with increasing coupling $\lambda$ -- this is also expected, since the
renormalization is directly caused by the electron-phonon coupling, see
Fig. \ref{fig2}(b). 

This renormalization is a direct
illustration of the general result of Eq. (\ref{redis}): the
electron-phonon coupling changes not only the properties of the
polaron (its effective mass), but also the disorder potential it
experiences.

\begin{figure}[t]
\includegraphics[width=0.9\columnwidth]{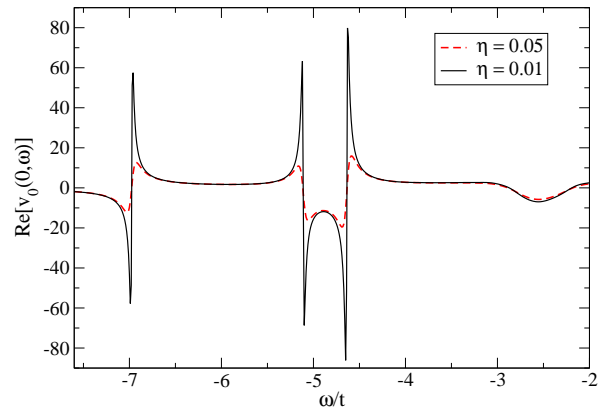}
 \caption{(color online) Real part of the additional on-site MA$^{(0)}$ disorder potential
 $v_0(0,\omega)$ over a wide energy range, for $U=2t, \Omega=2t,
 \lambda=0.8$ and two values of $\eta$.}
 \label{fig6}
 \end{figure}

However, it is very wrong to expect that the potential renormalization
can always be described by a simple rescaling by some overall
value. Indeed, Eq. (\ref{redis}) shows that the renormalized potential
is expected to be a function of energy, because of retardation
effects. This function is not roughly constant, instead it has
significant and very non-trivial dependence of $\omega$, as
illustrated by plotting $v_0(0,\omega)$ over a large energy range, in
Fig. \ref{fig6}. Similar curves are found for other values of the
parameters. Over a narrow range of energies around $\omega\sim -7.5t$
where the bound state forms for these parameters (see
Fig. \ref{fig3}), $v_0(0,\omega)$ varies slowly and can be
approximated as an overall negative constant. This explains why here
we can approximate $\epsilon^*_0(\omega) = -U + v_0(0,\omega) \approx
-U^*$, with $U^*>U$, as discussed above. At higher energies, however,
$v_0(0,\omega)$ goes through singularities and changes sign from
negative to positive and back. Although at first sight these
singularities are surprising, they should be expected based on
Eq. (\ref{v0}). The self-energy of the Holstein polaron is known to
have such singularities, especially at medium and higher couplings
where an additional second bound state forms and the continuum above
shows strong resonances spaced by $\Omega$. In particular, as
$\lambda \rightarrow \infty$ and the spectrum evolves towards the
discrete Lang-Firsov limit $E_n=-{g^2\over \Omega} + n\Omega$, the
self-energy has a singularity at the top of each corresponding
band. The renormalized potential of Eq. (\ref{v0}) is the difference
between two such curves,  displaced from each other. It is
thus not a surprise that it has such nontrivial behavior.

Physically, such strong retardation effects are not surprising, either, since
the additional potential $v_0(i,\omega)$ describes the 
scattering of the electron in the presence of the phonon
cloud. The structure of the phonon cloud varies 
with energy, for instance one expects quite different clouds
within the polaron band vs. at higher energies, in the continuum
of incoherent states with finite lifetime. This 
suggests that the diagrams of Fig. \ref{fig2}(b) that contribute most
to the series change with energy, and so does the total result. As a
final note, we mention that at these higher energies, MA$^{(1)}$
should be used. It is well-known that MA$^{(0)}$ fails to describe
properly the location of the polaron+one phonon continuum, since it does not include
the needed variational states.\cite{bar} This problem is
fixed at the MA$^{(1)}$ and higher levels.\cite{MA1}

To summarize, for this simple impurity problem the MA approximation is
found to agree well with results from DMC in describing the trapping
of the polaron. Although quantitatively not as accurate, besides
efficiency its main advantage is that the analytic equations that
describe MA allow us to understand the relevant physics. In
particular, we showed that coupling to bosons renormalizes the
disorder in a very non-trivial way.          
  
 \subsection{Impurity changing the electron-phonon coupling}
 \label{g-disorder}
 
We now assume that the impurity does not change the on-site energy,
but instead it modifies the value of electron-phonon coupling at
its site:
\begin{equation}
{\cal H}=\hat{T}+\Omega\sum_j b^{\dagger}_jb_j+\sum_jg_j
c^{\dagger}_jc_j(b_j+b^{\dagger}_j),
\end{equation} 
where $g_j=g+(g_d-g)\delta_{j0}$. Thus, $g_d$ and $g$ are the
electron-phonon couplings at the impurity site and in the bulk of the system,
respectively. Since $\hat{V}_d=0$ in this case, the non-interacting
part of the Hamiltonian is ${\cal{H}}_d={\cal{H}}_0$. Thus,
$G^d_{jj}(\omega-n\Omega)\rightarrow G^{(0)}_{jj}(\omega-n\Omega)\equiv
g_0(\omega-n\Omega)$ in the continued fractions, Eq.  (\ref{An}),
whose dependance on the site index $j$ is now through the coupling
$g_j$ only. As a result, the effective disorder potential
$v_0(j,\omega)$  vanishes everywhere except at the impurity site, $j=0$:
\begin{equation}
\label{Delta-g}
v_0(j,\omega)\equiv\Delta(\omega)\delta_{j0},
\end{equation}
where $\Delta(\omega) = g_d A_1(0,\omega) - g A_1(\omega)$, and
$A_1(0,\omega)$ is like in Eq. (\ref{Ac})  but with $g \rightarrow
g_d$. This shows that even though $\epsilon_i=0$ in this
case, the inhomogeneity gives rise to an effective potential
$\epsilon^*_i(\omega) = \delta_{i,0} \Delta(\omega)$. This is now local
because only when the phonon cloud is at the impurity site it can experience
the different coupling. 

Equation (\ref{eq:G-v0}) can now be solved analytically to find:
\begin{equation}
\EqLabel{sd}
\rho(0,\omega)=-\frac{1}{\pi}\mbox{Im}\left(
\frac{g_0(\tilde{\omega})}{1-\Delta(\omega)g_0(\tilde{\omega})}\right).
\end{equation}

 \begin{figure}
  \includegraphics[width=0.9\columnwidth]{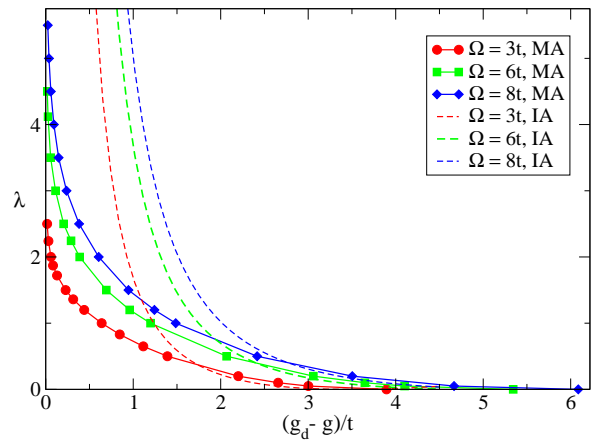}
 \caption{(color online) Phase diagram separating the regime where the
   polaron is delocalized (below the line) and trapped (above the
   line), as a function of the difference between the impurity and the
   bulk electron-phonon coupling, $g_d-g$. Symbols show MA$^{(0)}$
   results, while the dashed black lines correspond to the 
   instantaneous approximation.}
 \label{fig7}
 \end{figure}
 
We now find the critical values $g_d-g$ when an impurity state
emerges below the continuum, for given values of $g,\Omega$.  The
results are shown in Fig. \ref{fig7} for $g_d > g$, when the polaron
formation energy at the impurity site, $-g_d^2/\Omega$, is lower than
the bulk value $-g^2/\Omega$, and a bound state may be expected to
form even within the instanteneous approximation. Symbols show
MA$^{(0)}$ results, while the dashed lines are for the instantaneous
approximation. The two agree quantitatively only in the limit $\lambda
\rightarrow 0$.  This proves that the additional renormalization
included in MA is significant for this type of impurity, as well. We
note that all critical lines intercept the $x$-axis at a finite value,
{\em i.e.} for any value of $\Omega$ and $g=0$, there is a critical
finite value $g_d$ above which an impurity state forms. For example,
for $\Omega=3t$ this critical value is $g_d\approx 3.9t$. Its value
increases with increasing $\Omega$, as expected.

Unlike for an  impurity which changes the on-site  potential, and
which can bind at most one impurity state,
impurities which change the electron-phonon coupling can bind multiple impurity
levels. As $g_d$ increases and the energy of
the impurity level moves towards lower energies, additional bound
states, spaced by roughly $\Omega$, emerge whenever the distance
between the last one and the bulk polaron band is of order
$\Omega$. 

The origin of this sequence of bound states is straightforward to
understand in the limit $g_d\gg g, t$, where, the
Hamiltonian is, to zero order:
\begin{equation}
\nonumber
{\cal H} \approx g_dc^{\dagger}_0c_0(b^{\dagger}_0+b_0)+\Omega b^{\dagger}_0b_0,
\end{equation}
with $c^{\dagger}_0c_0\approx 1$ because the weight of the bound state is
concentrated at the impurity site. This Hamiltonian can be exactly
diagonalized with  the Lang-Firsov transformation\cite{lang-firsov}
and predicts a  series of equally spaced eigenenergies
$n\Omega-g_d^2/\Omega$.  For finite $t,g$, all  states which
lie below the bulk polaron continuum become impurity states, and
basically describe excited bound states with additional phonons at
the impurity site.

\begin{figure}[t]
  \includegraphics[width=0.9\columnwidth]{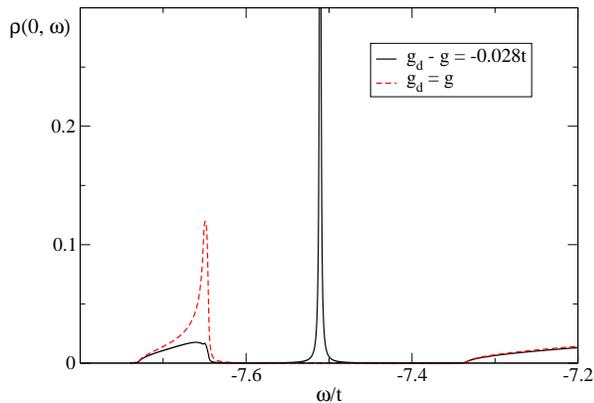}
 \caption{For large $\lambda$, in the clean system (dashed red line)
   the first polaron band  is 
   separated  by an energy gap from the next  features in the
   spectrum (here, the band associated with the second bound
   state). For $g_d < g$, an ``anti-bound'' impurity state is
   pushed inside this gap (black full line).  Parameters are $\Omega=t=1$,
   $\lambda=1.2$ and $\eta=10^{-3}$.}
 \label{fig8}
 \end{figure}

So far we have considered $g_d>g$, where a ground-state impurity level
can emerge. It is important to note that discrete peaks can also
appear for $g_d <g$, although not at low energies. This happens when
$\lambda$ is sufficiently large that there is a gap between the bulk
polaron band and higher features in the spectrum, such as the
polaron+one phonon continuum, or the band associated with the second
bound state, once it forms.\cite{bonca} A typical example is shown in
Fig. \ref{fig8}, where in the presence of an impurity with a weaker
coupling $g_d < g$ (full line), a discrete state appears above the
polaron band. Since most of its weight is removed from the bulk
polaron band (dashed line), we interpret this as being an
``anti-bound'' polaron state. A similar state is also expected to
appear for a repulsive on-site $U<0$ potential.
  
\subsection{Isotope impurity}  
 
The last case we consider in detail is an isotope impurity. Because of
its different mass $M_d\ne M$, both its phonon frequency $\Omega_d \sim
M_d^{-{1\over 2}}$, and its electron-phonon coupling $g_d\propto
1/\sqrt{M_d\Omega_d}\sim M_d^{-{1\over 4}}$, are changed. Interestingly,
both the effective coupling $\lambda_d =
g_d^2/(6t\Omega_d)=g^2/(6t\Omega)=\lambda$ and the polaron formation
energy $-g_d^2/\Omega_d = -g^2/\Omega$ show no isotope effect.\cite{iso}
As a result, within the instantaneous approximation of
Eq. (\ref{inst}) one would predict that the isotope is
``invisible'' and the polaron spectrum is basically unaffected by its presence.

We consider a single isotope impurity located at the origin:      
\begin{equation}
{\cal H}=\hat{T}+\sum_j
g_jc^{\dagger}_jc_j(b_j+b^{\dagger}_j)+\sum_j\Omega_j
b^{\dagger}_jb_j, 
\end{equation} 
where $\Omega_j=\Omega+(\Omega_d-\Omega)\delta_{j0}$ and
$g_j=g+(g_d-g)\delta_{j0}$ are chosen such that $\lambda_d = \lambda$.

Just like in the previous section, because there is no on-site
disorder $\epsilon_i=0$, we have $G^d_{jj}(\omega) =
G^{(0)}_{jj}(\omega)$ and the effective disorder potential is again
local, {\em i.e.} it vanishes everywhere but at the impurity site. As
a result, the LDOS at the impurity site is given by Eq. (\ref{sd}),
except that here:
\begin{equation}
\label{Delta-OMEGA}
\Delta(\omega)=\Sigma_d(\omega) -
\Sigma_\mathrm{MA^{(0)}}(\omega),
\end{equation}
where $\Sigma_d(\omega)$ has the same functional form like
$\Sigma_\mathrm{MA^{(0)}}(\omega)$, but with $g\rightarrow g_d,
\Omega \rightarrow \Omega_d$. 

From investigations of this LDOS at different parameters we find  that
there exists a threshold value of the effective coupling, $\lambda^*$, below 
which  low-energy bound states do not
form irrespective of how small or large $M_d/M$ is. In other words,
for $\lambda < \lambda^*$, the behavior agrees with the prediction of
the instantaneous approximation.

We can estimate a bound on $\lambda^*$ as follows. Consider the case
of a  very light isotope, so that  $\Omega_d,g_d\gg \Omega, g, t$. In
this limit,  $\Sigma_d(\omega) \rightarrow
-g_d^2/\Omega_d=-6t\lambda$. The bound state appears when the LDOS is
singular because its denominator vanishes: 
\begin{equation}
\label{peak-OMEGA}
1-\Delta(\omega)g_0(\tilde{\omega})=0\rightarrow\Sigma_\mathrm{MA^{(0)}}(\omega)
+ 1/g_0(\tilde{\omega}) = -6t\lambda,
\end{equation}
after using Eq. (\ref{Delta-OMEGA}) and $\Sigma_d(\omega)\approx -6t\lambda$.

\begin{figure}[t]
  \includegraphics[width=0.9\columnwidth]{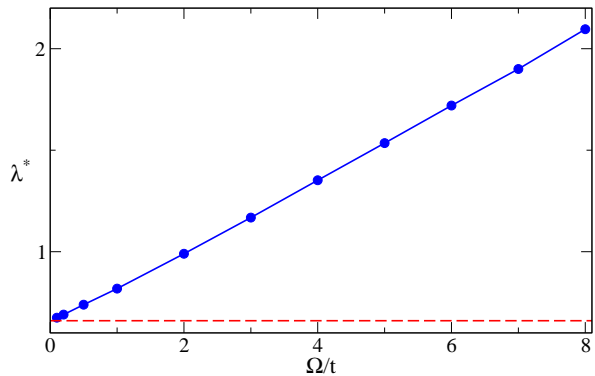}
 \caption{(color online) Critical effective coupling $\lambda^*$ above which an
   impurity state may appear for a sufficiently light isotope. Below
   this line, polarons cannot be bound near isotopes. Symbols shows
   MA$^{(0)}$ results. The dashed line is the analytic low-bound for
   $\lambda^*$   discussed in the text.} 
 \label{fig9}
 \end{figure}

Consider now the limiting case when a bound state emerges just below
the  bulk polaron  ground state, {\em i.e.} Eq. (\ref{peak-OMEGA})
has a solution at 
$\omega\leq\varepsilon^{pol}_{gs}$. In the clean system, the polaron
ground-state energy $\varepsilon^{pol}_{gs}$ is the lowest pole of $G(k=0,\omega) = [\omega -
\epsilon_{{\bf k}=0}-\Sigma_\mathrm{MA^{(0)}}(\omega)]^{-1}$, so it satisfies:
$\varepsilon^{pol}_{gs} = -6t +
\Sigma_\mathrm{MA^{(0)}}(\varepsilon^{pol}_{gs} )$. Using this
in Eq. (\ref{peak-OMEGA}) suggests that a solution can exist if
$\lambda > \lambda^*$, where
\begin{equation}
\label{lambda-c}
\lambda^*=\left|
\frac{\varepsilon^{pol}_{gs}}{6t}\right|-1 - {1\over 6t g_0(-6t)}, 
\end{equation}   
with $g_0(-6t)\approx -1/3.96t$. Since
$\varepsilon^{pol}_{gs}\rightarrow -6t$ as $\lambda\rightarrow
  0$, we expect that $\lambda^* \rightarrow 0.66$ in this limit, and
  that it increases as $\varepsilon^{pol}_{gs}$ becomes more negative,
  for example  with increasing $\lambda$. These considerations are
  confirmed by the data shown in Fig. \ref{fig9}. Here, the symbols
  show values of $\lambda^*$ found numerically with MA$^{(0)}$, and
  the dashed line is the lower bound of 0.66, discussed above. 
  
\begin{figure}[b]
  \includegraphics[width=0.9\columnwidth]{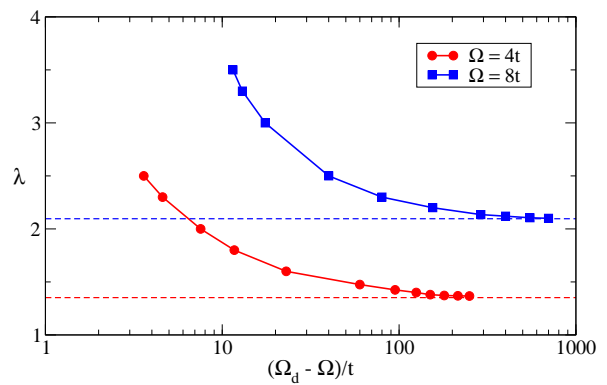}
 \caption{(color online) Phase diagram separating the regime where the
   polaron is delocalized (below the line) and trapped (above the
   line), as a function of the difference between the
   $\Omega_d-\Omega$, on a logarithmic scale. Symbols show  MA$^{(0)}$
   results for 
   $\Omega=4t, 8t$. As $\Omega_d\rightarrow \infty$, these critical lines
   converge towards their corresponding $\lambda^*$ (dashed lines),
   below which polaron
  states are always delocalized.} 
 \label{fig10}
\end{figure}

For $\lambda > \lambda^*$, bound impurity states can appear near
isotopes if $g_d$ and $\Omega_d=g_d^2/(6t\lambda)$ are sufficiently
large. In Fig. \ref{fig10} we show critical lines for two cases,
$\Omega=4t, 8t$. The symbols show the MA$^{(0)}$ results, which
converge towards their corresponding $\lambda^*$ values as $\Omega_d
\rightarrow \infty$, as expected. Of course, the largest values
considered for $\Omega_d$ are unphysical; we use them only to
illustrate the convergence towards $\lambda^*$.

\begin{figure}[t]
  \includegraphics[width=0.9\columnwidth]{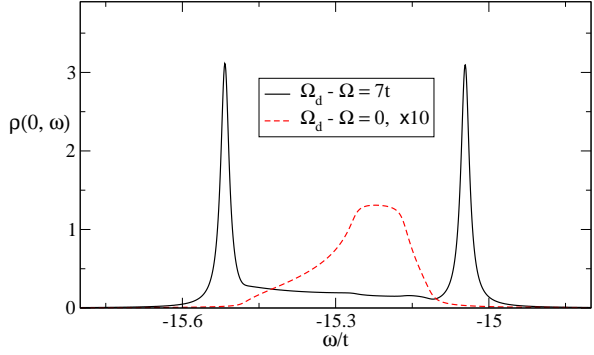}
 \caption{(color online) LDOS at the impurity site near an isotope with
   $\Omega_d=\Omega+7t$ (full line). Two discrete states, one above
   and one below the bulk polaron band, are seen. For comparison, the
   DOS in the clean system (multiplied by 10) is shown as a dashed line.
Parameters are
$\Omega=4t$, $\lambda=2.5$ and $\eta=10^{-2}$.} 
 \label{fig11}
 \end{figure}

The existence of a region of the parameter space where bound polaron
states appear near an isotope is in direct contradiction of the
instantaneous approximation, and again illustrates the importance of
the renormalized disorder potential $\Delta(\omega)$ which makes their
trapping possible. In this context, it is worth mentioning that there
is clear evidence for electronic states bound near isotope O$^{16}$
defects in CuO$_2$ planes, \cite{bernhard} although the precise nature
of these states has not been clarified and the measurements are certainly not in
the extremely underdoped regime where our single-polaron results are valid.

Interestingly, when such bound states form near an isotope, the
spectrum is different than that for the other two types of
impurities. As shown in Fig. \ref{fig11}, bound states now appear
simultaneously both below and above the bulk polaron continuum, not
just below it. This provides a possible ``fingerprint'' for polarons
trapped near isotopes. Finally, we note that even when no low-energy
impurity state is observed, it is again possible to have higher energy bound
states inside the gaps opening between various features in the bulk
polaron spectrum.

To summarize, in the presence of isotope defects, polarons in the weakly
coupled regime $\lambda<\lambda^*$ always remain delocalized. Only for
$\lambda>\lambda^*$ it is possible to trap polarons near an
isotope. This makes this case quite distinct from the other two cases,
where bound states exist for any $\lambda$ if the impurity is strong
enough.

\section{Summary and conclusions}

We studied the threshold for the emergence of polaron bound states
near various types of single impurities, using the momentum-average
approximation. Electron-phonon coupling was shown to strongly
renormalize the impurity potential in a nontrivial way which includes
strong retardation effects. This is a feature that is completely
absent in the instantaneous approximation, which is the only other
available ``simple'' description of this problem.

We considered the simplest models of impurities that change the
strength of the on-site 
energy, the local electron-phonon coupling, or are isotope
substitutions that modify both the coupling and the phonon energy. We
calculated the polaron binding phase diagrams for each case. The first
case had been considered previously by numerical
methods,\cite{mishchenko,hague} and our results are in good
quantitative agreement with their predictions. To our knowledge, the
other two cases have not been investigated before.  We showed that in
the first two cases bound states always exist for a sufficiently
strong impurity, however the polaron remains delocalized for the case
of isotope substitution of arbitrary strength if the effective
coupling is weaker than a threshold value, $\lambda^*$. Differences in
the LDOS at the impurity site have also been found, such as the possibility
to bind multiple states near an impurity that changes the coupling, or
the unusual fingerprint of discrete states both below and above the
bulk polaron continuum, for an isotope bound state.

Of course, a realistic description of an impurity in a real system may
combine several of these inhomogeneities, and even the form of the
electron-phonon coupling could be affected. MA gives an efficient yet
quite accurate way to deal with such cases, and can be easily
generalized to other types of couplings where MA has been used
succesfully to describe bulk properties.

Whereas we expect the single impurity results to remain valid for a
system with multiple impurities if the mean free path is long and the
polaron interacts with one impurity at a time, in the presence of
significant disorder, when multiple scattering processes become
important, the polarons can undergo Anderson localization. This limit
has been addressed within a generalized DMFT,\cite{fehske} however we
believe that our simpler formulation might provided additional insight
and uncover previously unexplored aspects of Anderson localization for
polarons. Such work is currently in progress.

\begin{acknowledgments}
 We thank Andrey Mishchenko and Holger
Fehske for useful discussions and sharing their results. This work was
supported by NSERC and 
CIFAR.
\end{acknowledgments}

\appendix
\section{IMA$^{(1)}$}

At the IMA$^{(1)}$ level, one also allows processes in which one phonon is
away from the phonon cloud. These are described by the propagators
$S_n(i,l,j;\omega)=\langle 0| c_i G(\omega) c^{\dagger}_l b^{\dagger
  n-1}_l b^{\dagger }_j | 0\rangle$ with $j\ne l$. In terms of these,
Eq. (\ref{eq:Dyson}) 
can be written as  
\begin{equation}
G_{ij}(\omega)=G_{ij}^d(\omega)+\sum_l g_l S_1(i,l,l;\omega)G_{lj}^d(\omega).
\label{G-S1}
\end{equation}	 

Once again we apply the Dyson identity to $S_1$
\begin{eqnarray}
\label{S1}
S_1(i,l,j;\omega)&=&g_jG_{jl}^d(\omega-\Omega_j)G_{ij}(\omega) \\
	&& + \sum_m g_mG_{ml}^d(\omega-\Omega_j)S_2(i,m,j;\omega). \nonumber
\end{eqnarray}	

This exact equation relates $S_1$ to the  propagators $S_2$. We can similarly
find the equation of motion of all  higher, $n\geq 2$,
$S_n(i,l,j;\omega)$ with $l \neq j$ and $l=j$, separately. For $l \neq
j$ we have 
\begin{eqnarray}
S_n(i,l,j;\omega)&&=g_lG_{ll}^d(\omega-(n-1)\Omega_l-\Omega_j)\\
	&& \times[(n-1)S_{n-1}(i,l,j;\omega)+S_{n+1}(i,l,j;\omega)],\nonumber
\end{eqnarray}
where we now ignore contributions from terms with a second phonon away from the
polaron cloud, as they are exponentially smaller than those we
kept.
This admits the solution
$S_n(i,l,j;\omega)=B_n(l,j;\omega)S_{n-1}(i,l,j;\omega)$, where  
\begin{eqnarray}
 B_n(l,j;\omega)&=&\frac{(n-1)g_lG_{ll}^d(\omega-(n-1)\Omega_l-\Omega_j)}
 {1-g_lG_{ll}^d(\omega-(n-1)\Omega_l-\Omega_j)B_{n+1}(l,j;\omega)}\nonumber 
 \\  
 &=& A_{n-1}(l,\omega-\Omega_j). 
\end{eqnarray}

For $l=j$ and $n\geq 2$, 
$S_n(i,l,l;\omega)=F_{il}^{(n)}(\omega)$ and we get the same
solution as in MA$^{(0)}$,
i.e. $S_n(i,l,l;\omega)=A_n(l,\omega)S_{n-1}(i,l,l;\omega)$. The
relations between $S_2$ and $S_1$ are used in 
Eq. (\ref{S1}) to turn it into an equation between $S_1(\omega)$ and
$G_{ij}(\omega)$ only  
\begin{multline}
S_1(i,l,j;\omega)=g_jG_{jl}^d(\omega-\Omega_j)G_{ij}(\omega) \\
	 +
g_jG_{jl}^d(\omega-\Omega_j)A_2(j,\omega)S_{1}(i,j,j;\omega) \nonumber
\\  
	 + \sum_{m\neq j} g_mG_{ml}^d(\omega-\Omega_j)
A_1(m,\omega-\Omega_j)   S_{1}(i,m,j;\omega). \nonumber
\end{multline}
Together with Eq. (\ref{G-S1}), this can be solved to find
$G_{ij}(\omega)$. However, it is again convenient to explicitly
extract the ``average'' contributions, to make these equations more efficient.

We therefore remove the homogeneous part from Eq. (\ref{S1}) and
include it into a renormalized energy:
\begin{widetext}
\begin{multline}
\label{ugly}
S_1(i,l,j;\omega)=g_jG_{jl}^d(\bar{\omega}_j)[G_{ij}(\omega)+(A_2(j,\omega)-
  A_1(j,\omega-\Omega_j))S_1(i,j,j;\omega)]\\
+ \sum_{m\neq j} g_mG_{ml}^d(\bar{\omega}_j)
[A_1(m,\omega-\Omega_j)-A_1(\omega-\Omega)] S_{1}(i,m,j;\omega).
\end{multline}
where $\bar{\omega}_j=\omega-\Omega_j-gA_1(\omega-\Omega)$, and
$A_1(\omega-\Omega)$ is given by Eq. (\ref{Ac}) for the
``average'' clean system.

The sum on the rhs of Eq. (\ref{ugly}) again convergences for a very
small cutoff, only sites $m$ very close to $j$ need to be
included. Its general solution is of the form:
\begin{equation}
\EqLabel{Se}
S_1(i,l,j;\omega)= x_{jl}(\omega) [G_{ij}(\omega)+(A_2(j,\omega)-
  A_1(j,\omega-\Omega_j))S_1(i,j,j;\omega)], 
\end{equation}
where
$$
x_{jl}(\omega) = g_jG_{jl}^d(\bar{\omega}_j) + \sum_{m\neq j}
g_mG_{ml}^d(\bar{\omega}_j) 
[A_1(m,\omega-\Omega_j)-A_1(\omega-\Omega)] x_{jm}(\omega).
$$ 
\end{widetext} 
In fact, using $x_{jl}(\omega) = g_jG_{jl}^d(\bar{\omega}_j)$ is
already a very good approximation, since the terms in the sum are
exponentially small -- but one can go beyond this. Once 
$x_{jj}(\omega)$ is known, from Eq. (\ref{Se}) we find
$S_1(i,j,j,\omega) = \Lambda_j(\omega) G_{ij}(\omega)$, where
$$
\Lambda_j(\omega) =
\frac{x_{jj}(\omega)}{1-x_{jj}(\omega)\left(A_2(j,\omega)-
  A_1(j,\omega-\Omega_j)\right) }.
$$
This can now be used in  Eq. (\ref{G-S1}) to turn it
into an equation for $G_{ij}(\omega)$ only: 
\begin{equation}
G_{ij}(\omega)=G_{ij}^d(\omega)+\sum_l g_l \Lambda_l(\omega)
G_{il}(\omega)G_{lj}^d(\omega). 
\end{equation}

As we did in Eq. (\ref{eq:G-v0}) for MA$^{(0)}$, this can be made
efficient to solve by subtracting the MA$^{(1)}$ self-energy and
including it into the energy argument 
\begin{equation}
G_{ij}(\omega)=G_{ij}^d(\tilde{\omega})+\sum_l G_{il}(\omega)
v_1(l,\omega)G_{lj}^d(\tilde{\omega}), 
\end{equation}
in which $\tilde{\omega}=\omega-\Sigma_{\mathrm{MA^{(1)}}}(\omega)$
and
$v_1(l,\omega)=g_l\Lambda_l(\omega)-\Sigma_{\mathrm{MA^{(1)}}}(\omega)$. Here,
$\Sigma_{\mathrm{MA^{(1)}}}(\omega)$ is the value of
$g_l\Lambda_l(\omega)$ in the clean, ``average'' system:
$$
\Sigma_{\mathrm{MA^{(1)}}}(\omega) = \frac{g^2
  g_0(\bar{\omega})}{ 1- g
  g_0(\bar{\omega})
  \left(A_2(\omega)-A_1(\omega-\Omega)\right)} 
$$
where now $\bar{\omega} = \omega-\Omega-gA_1(\omega-\Omega)$.\cite{MA1}
This
completes the calculation of Green's function within inhomogeneous
MA$^{(1)}$ approximation.

\end{document}